\documentclass[twocolumn,showpacs,preprintnumbers,amsmath,amssymb]{revtex4}
\usepackage{graphicx}
\usepackage{dcolumn}
\usepackage{bm}
\begin{document}

\title{Dynamical response function of a compressed lithium monolayer.}

\author{A. Rodriguez-Prieto$^{1,2}$}
\author{V. M. Silkin$^1$}
\author{A. Bergara$^{1,2}$}
\author{P. M. Echenique$^{1,3}$}

\affiliation{1 Donostia International Physics Center (DIPC), Paseo de Manuel Lardizabal, 20018, Donostia, Basque Country, Spain\\ 2 Materia Kondentsatuaren Fisika Saila, Zientzia eta Teknologia Fakultatea, Euskal Herriko 
Unibertsitatea, 644 Postakutxatila, 48080 Bilbo, Basque Country, Spain\\ 3 Materialen Fisika Saila, Kimika Fakultatea,\\ Euskal Herriko Unibertsitatea, 20080, Donostia, Basque Country, Spain}

\begin{abstract}
Since recent both theoretical and experimental results have proved that the simple behaviour light alkaline metals 
present at equilibrium breaks when high pressures are applied, they have become an important object of study in 
Condensed Matter Physics. On the other hand, development of new techniques in the atomic 
manipulation allows the growth of atomic monolayers (ML's), therefore rising the interest to analyze low dimensional 
systems under different conditions. In particular, new \textit{ab initio} calculations performed for a lithium ML 
show that its electronic properties experience important modifications under pressure, which could lead to 
significant modifications in its dynamical response function. 
In this article we perform \textit{ab initio} calculations of the dynamical response function of a 
lithium ML analyzing its evolution with increasing applied 
pressure. We show that besides the well known intraband and interband plasmons, 
rising electronic density induces characteristic features of acoustic plasmons 
related to the presence of two types of carriers at the Fermi level.
\end{abstract}

\maketitle

\section{Introduction}
Light alkali metals have lately become an important object of study due to the 
complexity induced by pressure in their behaviour, in contrast to the simple picture they present under normal 
conditions of pressure and 
temperature. At equilibrium, alkalies are frequently considered to be simple metals due to 
their crystallization in high symmetric 
structures, monovalency and high conductivity. At the same time the interaction between valence electrons and ionic cores, 
or pseudopotential, has been argued to be weak in such systems. 
As a consequence, the nearly free electron (NFE) model accurately describes their electronic properties in absence 
of applied 
pressure \cite{wigner} as it can be easily checked by their almost spherical Fermi surfaces. However, when high pressures are 
applied, the magnitude of the pseudopotential rises, the core electrons substantially overlap and the otherwise valid NFE model breaks. 
Neaton \textit{et al} \cite{NA} theoretically
 analyzed compressed bulk lithium, predicting that high pressures could induce phase transitions to less 
symmetric, lower coordinated structures, associated to electronic localizations. These theoretical predictions were 
experimentally confirmed by Hanfland \textit{et al} \cite{hanfland}, who found that lithium undergoes several phase transitions from a 
simple, high symmetric, $bcc$ structure at equilibrium to a complex $cI16$ with 16 atoms per unit cell at around 40 GPa. 
It is important to mention 
that pressure induced transitions from simple to more complex structures are not singular to lithium but have also 
been observed in heavier alkalines \cite{neaton}. Another important feature which shows up the complexity induced by pressure in the 
behaviour of light alkalies is that despite the superconducting transition for lithium at equilibrium has not been 
found yet, $T_{c}<100\,\mu$K \cite{liambient}, when compressed to around 30 GPa, $T_{c}$ rises up to 15 K \cite{shimizu,deemyad}, 
becoming the highest 
transition temperature between simple elements \cite{ashcroft_nature}. Therefore, the characterization and understanding of the physical properties of compressed light elements becomes a priority. On the other hand, the development of new techniques for the atomic manipulation allows the growth of atomic monolayers 
(ML's) on 
inert substratum, semiconductors or noble gases. These new possibilities rise the interest to analyze physical properties 
of low-dimensional systems under different conditions, as it could be the case of a lithium ML under pressure. In addition, 
extending our conclusions to the bulk will also give another perspective to understand the physical origin under the 
experimentally observed features in compressed lithium. Previous \textit{ab initio} analysis of the structural and electronic properties of a lithium ML \cite{aitor,prb} reveal important 
modifications in both its band structure and Fermi \textit{line}, which will also lead to significant modifications on its 
dynamical response function, up to now just studied at equilibrium \cite{aitor_plasmon}. In this article we perform 
calculations of  the dynamical response function of a lithium ML at different pressures, analyzing its evolution with increasing 
electronic density. Besides the common intraband and interband plasmons, at a certain value of the lateral 
pressure applied to the ML, we observe characteristic features of acoustic plasmons related to the presence, at this 
pressure range, of two types of carriers at the Fermi level. In Section \ref{sec:T} we describe the theoretical and 
computational background of this work. Results and Discussion are presented at Section \ref{sec:R}. Unless otherwise 
stated we use atomic units throughout, i.e., $e^{2}=\hbar=m_{e}=1$. 

\section{Theoretical and computational background.}
\label{sec:T}
We consider a perturbing positive charge located far from the ML, $z_{0}>>d$ ($d$ being the thickness of the ML). Then, the 
differential cross section for a process in which the electron is scattered with energy $\omega$ and momentum transfer 
$|\bf{q}|$ is proportional to Im $g(\bf{q},\omega)$, where $g(\bf{q};\omega)$ is defined as the density response 
function of the ML \cite{persson,tsuei},
\begin{eqnarray}
\label{eq:g}
&g({\bf{q}};\omega)=-\frac{2\pi}{|{\bf{q}}|}\int\,dz\int\,dz'\times&
\nonumber\\
&\times \chi_{{\bf{G}}=0,{\bf{G}'}=0}({\bf{q}},z,z';\omega)e^{|{\bf{q}}|(z+z')},&
\end{eqnarray}
where $\bf{q}$ belongs to the two-dimensional Brillouin zone (2DBZ) and ${\bf{G}}$ and ${\bf{G}'}$ are reciprocal two-dimensional lattice vectors. In order to obtain $g({\bf{q}};\omega)$ we strictly follow the method presented in Refs. \cite{aitor_plasmon,slava_plasmon}. Thus, our calculations are performed within the framework of time dependent density functional theory (TDDFT), where the random phase approximation (RPA) \cite{pines} has been applied. We use a local density 
approximation (LDA) in order to calculate the total energy and one-particle eigenvalues and eigenvectors 
of the ground state, which were evaluated with the use of Troullier-Martins Li pseudopotentials \cite{tm}. 
The description of the ML is implemented by a supercell which contains 20 layers of 
vacuum between ML's in order to minimize interaction between them. Effects of compression were simulated by reducing the 
lattice parameter. For the evaluation of $\chi^{0}_{{\bf{G}},\bf{G}'}$, previous step in the calculation of $\chi_{{\bf{G}},\bf{G}'}$, we implement up to $\sim$80 bands. In addition, 
we use $\sim7700$ $\bf{k}$-points for the sampling of the 2DBZ.

\section{Results and Discussion}
\label{sec:R}
\begin{figure}[h!]
\includegraphics[width=\linewidth,clip=true]{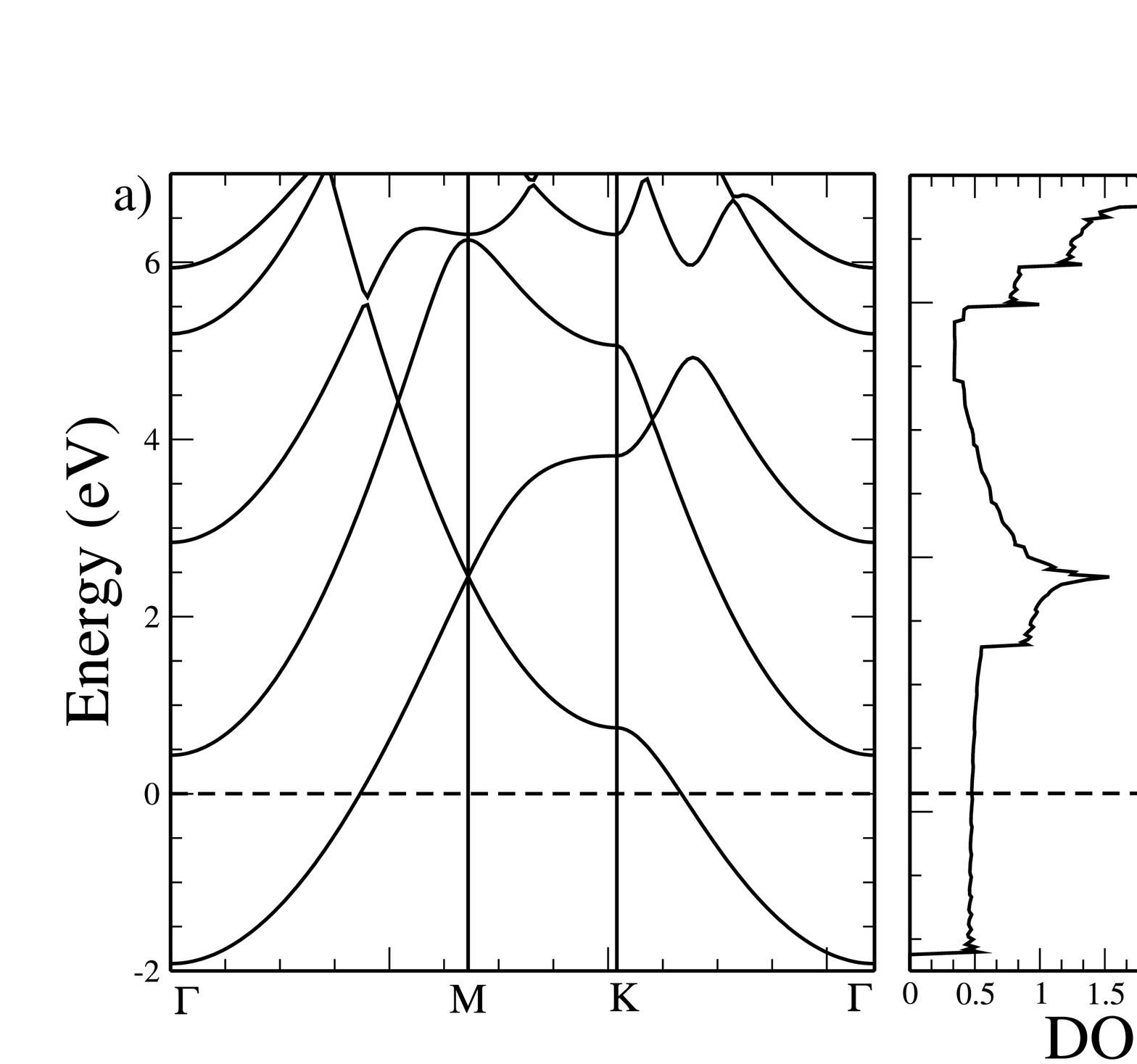}\\
\includegraphics[width=\linewidth,clip=true]{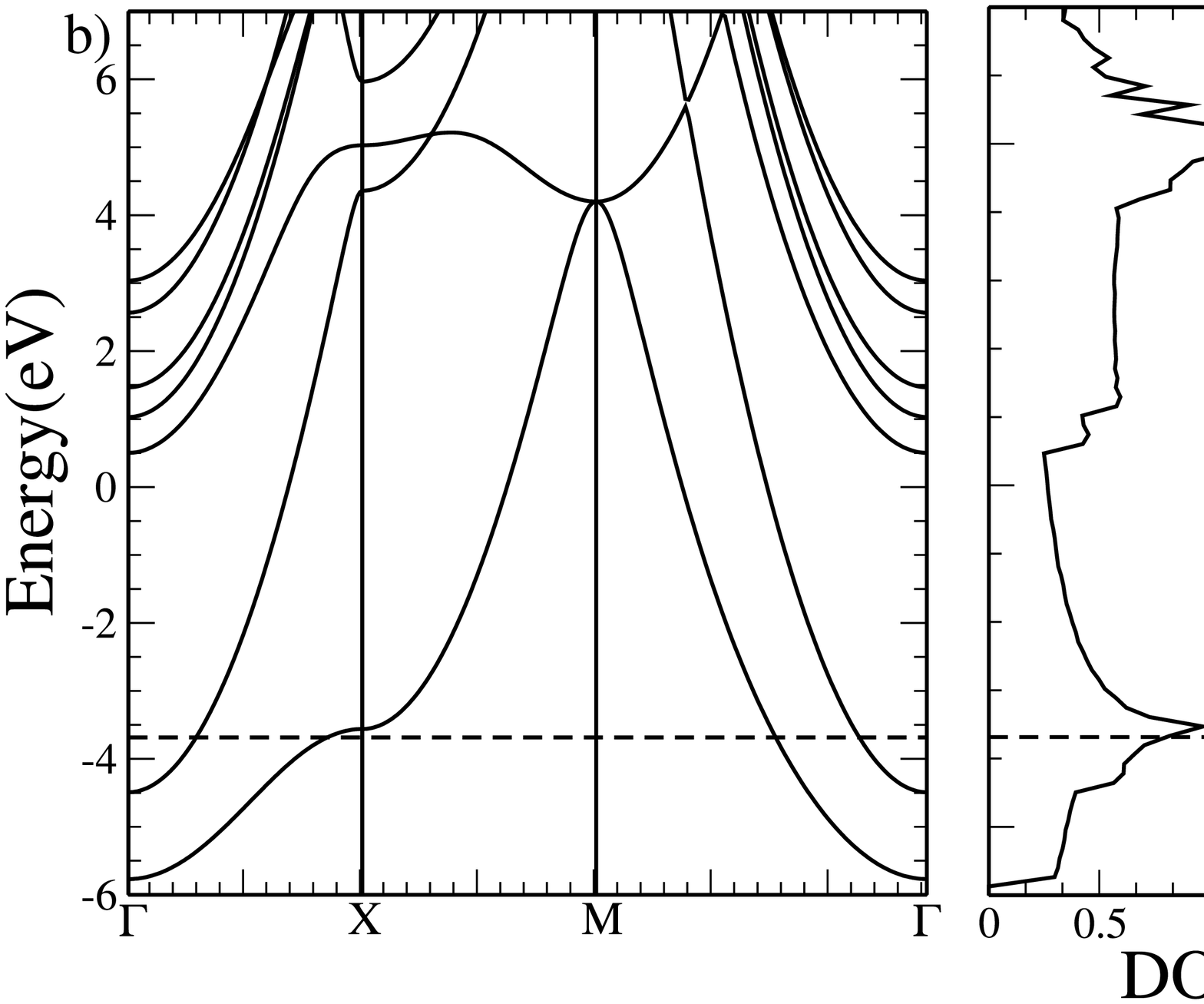}
\caption{Band structure and density of states (DOS) for (a) a hexagonal lithium ML at equilibrium ($r_s=3.02$ a.u.) and 
(b) a square structure at $r_{s}=2.15$ a.u. (close to the structural transition). The Fermi energy is represented by the 
dotted line. The bandwidth of the first band decreases with pressure and induces the occupation of the second band.}
\label{fig:bandas_monoat}
\end{figure}

\begin{figure}[h!]
\hspace*{0.5cm}\includegraphics[scale=0.28]{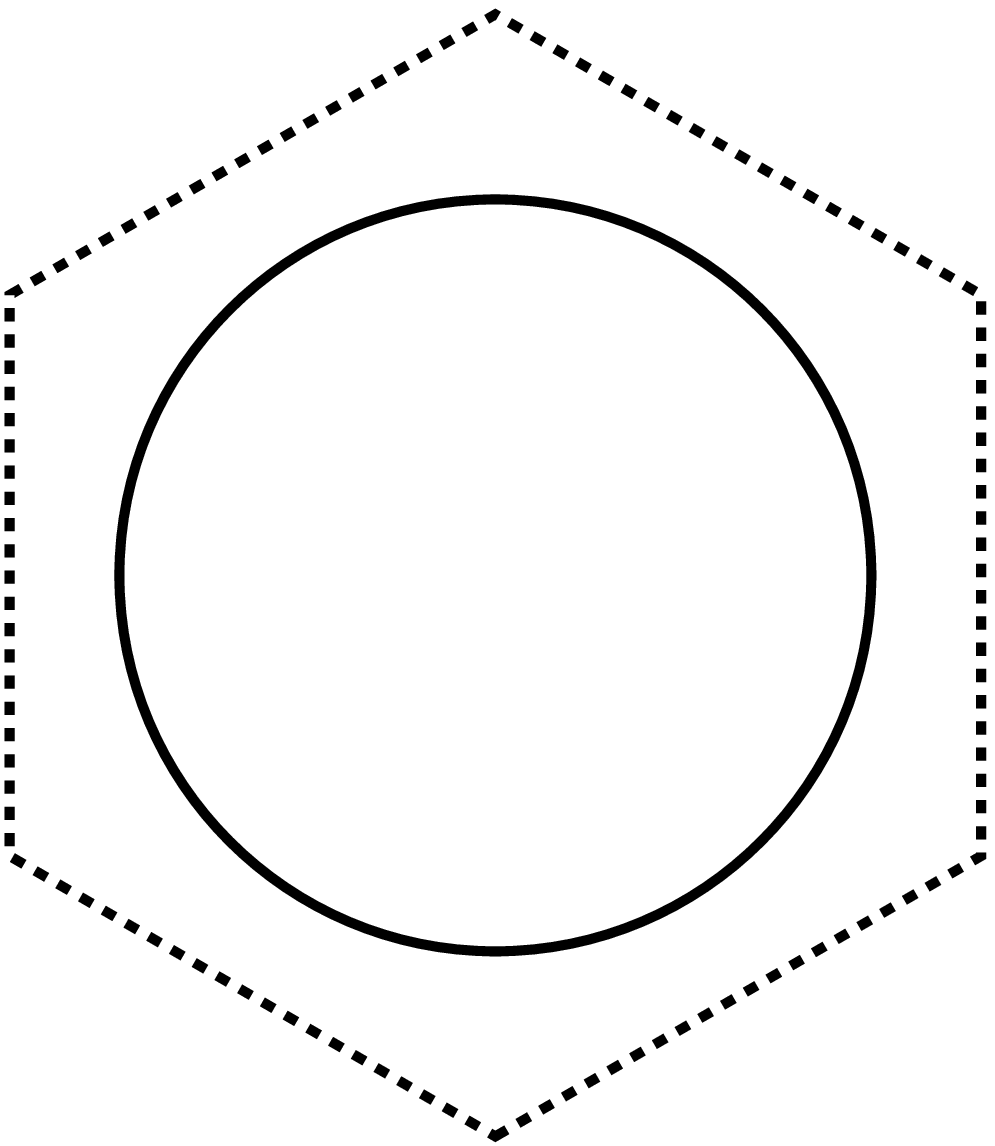}\hspace*{0.75cm}
\includegraphics[scale=0.28]{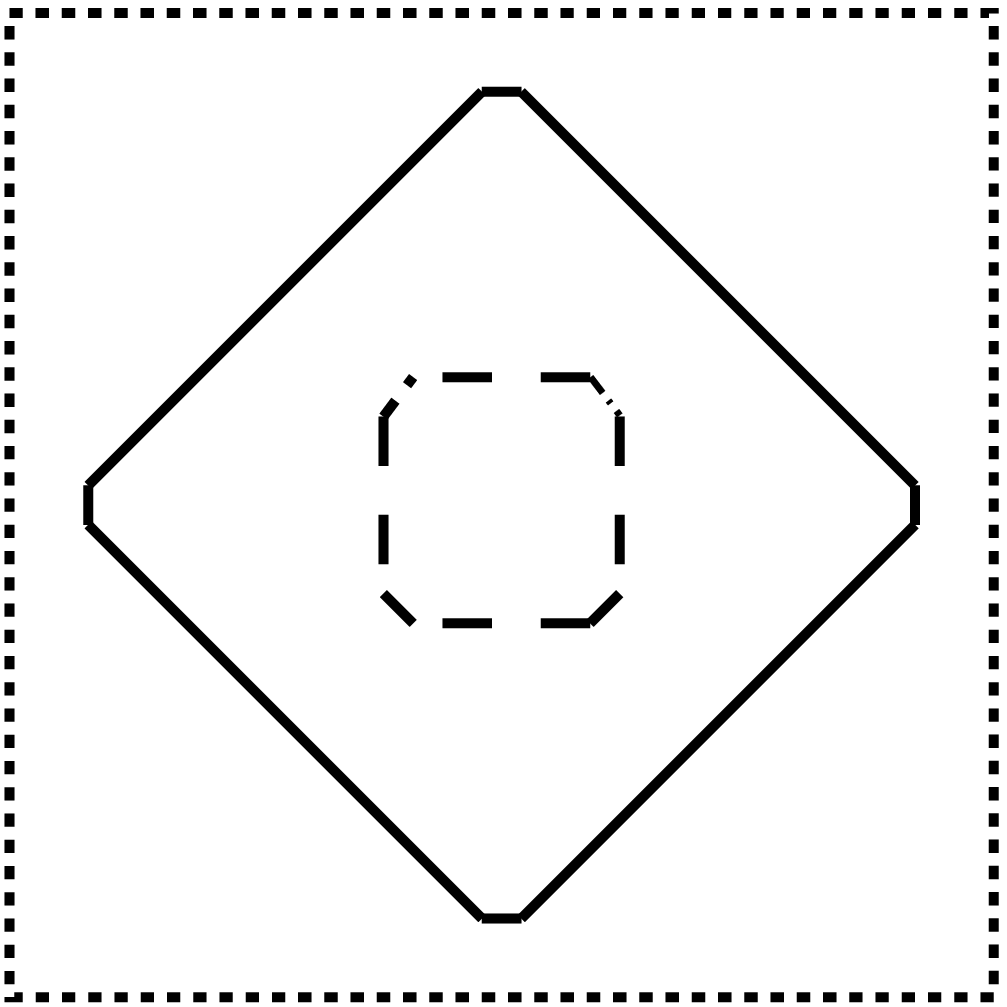}
\caption{Fermi \textit{lines} of a hexagonal lithium ML at equilibrium, $r_{s}=3.02$ (left), and a square lithium ML at $r_{s}=2.15$ (right). Solid line corresponds to the Fermi \textit{line} of the $s$-band. For values of the density $r_{s}\leq2.4$, the $p_{z}$-band starts to be occupied, so that this band is represented by a dashed line. The zone boundary is plotted with a dotted line.}
\label{fig:fermi}
\end{figure}

According to previous \textit{ab initio} calculations, the lithium ML adopts an hexagonal ($hex$) structure at 
equilibrium, which corresponds to $r_{s}=3.02$\footnote{$r_{s}$ is the two-dimensional linear density parameter, defined by the relation $A/N=\pi(r_{s}a_{0})^{2}$ where $N$ is the number of nuclei in a ML with area $A$ and $a_{0}$ is the Bohr radius.}. However, with applying pressure, a phase transition to the more opened 
square ($sq$) structure is produced at $r_{s}=2.25$. In Fig. \ref{fig:bandas_monoat} we plot the band structure and density of states (DOS) of a lithium ML corresponding to 
the $hex$ structure at equilibrium and the $sq$ structure with $r_{s}=2.15$, close to the structural phase transition. 
We can observe a quasi free electron-like behaviour at equilibrium, where only one band with a parabolic-type dispersion 
of $s$-character is occupied. In addition, the DOS for occupied states is almost 
constant, reflecting the quasi two-dimensional character of the ML. However, we can notice important modifications at 
higher pressures: the first band flattens associated to the increasing of the band gap at the zone boundary, indicating a 
clear electronic localization. At the same time, as a consequence of the decrease in the energy difference between the 
first two bands at $\Gamma$, the $p_{z}$-band, which is antisymmetric in the direction perpendicular to the plane of the 
ML, starts to be occupied. Therefore, we have two types of carriers at the Fermi energy, a fact which 
plays a fundamental role in the pressure induced properties of the electronic collective excitations, as will be explained 
bellow. It is noteworthy that the Fermi \textit{line} of the ML also suffers significant deviations with increasing electronic density \cite{prb}, as it is shown in Fig. \ref{fig:fermi}.

These important pressure induced modifications in both the band structure and the Fermi \textit{line} of the ML are expected
 to have a profound effect in the dynamical response function. In what follows, we will analyze the loss function, 
proportional to the imaginary part of $g({\bf{q}};\omega)$, at two different selected densities, $r_{s}=3.02$ (equilibrium) and 
$r_{s}=2.4$. At the latter density the phase transition has not yet been produced whereas the $p_{z}$-band is already 
occupied. In such a way, we can conclude that the physical origin of the interesting phenomena which will be described bellow is not related to the $hex\rightarrow sq$ phase transition, but is a direct consequence of the presence of two different conductors at the Fermi energy.
\begin{figure}[h!]
\includegraphics[width=\linewidth,clip=true]{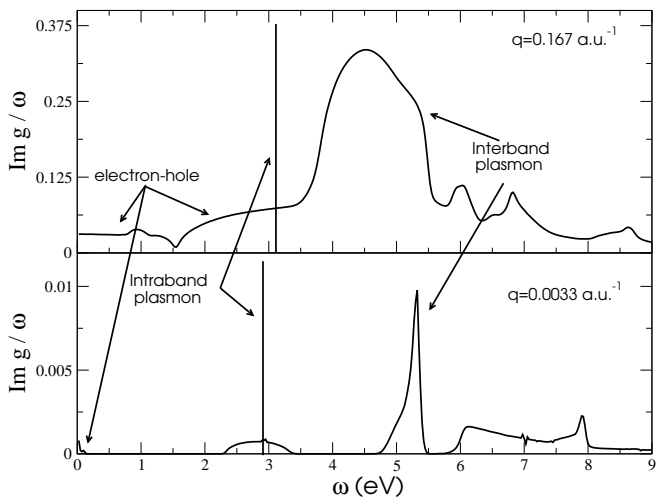}
\caption{Loss function Im $g({\bf{q}};\omega)$ of the hexagonal lithium ML at equilibrium for two different values of the wavefunction in the $\Gamma K$ direction. One can see one-particle electron-hole excitations and two peaks associated to the 
intraband and interband plasmons. }
\label{fig:eq_plasmon}
\end{figure}
In Fig. \ref{fig:eq_plasmon} we present the loss function of a lithium ML at equilibrium for two selected wavevectors in 
the $\Gamma K$ direction. As we are mainly interested in collective excitations, small momentum transfers are considered. 
Our spectra consists in one-particle electron-hole excitations with features corresponding to intraband excitations 
at small energies and interband excitations at higher energies, together with two plasmon peaks. The first 
peak is situated at an energy close to $\sim$3 eV and linked to the intraband collective excitations 
inside the $s$-band, whereas the second one with energy of $\sim$5 eV is related to the interband plasmon, which physically 
corresponds to collective motion of electrons in the direction perpendicular to the ML.
\begin{figure}[h!]
\vspace*{0.5cm}
\hspace*{1.0cm}\includegraphics[scale=0.23]{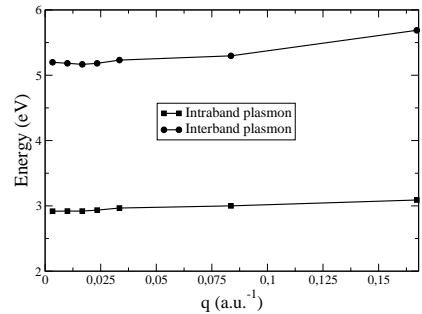}
\caption{Dispersion curves of the \textit{ab initio} collective intraband (squares) and interband (circles) excitations in the $\Gamma K$ direction. The non-zero minimum energy in the limit ${\bf{q}}\rightarrow 0$ indicates that electrons located at neighbouring layers feel the interaction.}
\label{fig:eq_dispersion}
\end{figure}
Dispersion curves for both intraband and interband collective excitations are plotted in Fig. \ref{fig:eq_dispersion}. 
As our selected wavefunctions are much smaller than the characteristic Fermi momentum of the ML, 
we would expect a $\sqrt{\bf{q}}$ dispersion \cite{stern} for the two-dimensional intraband plasmon, instead of the non-zero minimum energy for 3D plasmons \cite{pines}. However, the Fourier transform of the electronic potential of 
electrons located at neighboring layers decays exponentially with $|{\bf{q}|}L$, so that when dealing 
with vectors whose size is  
comparable to the inverse of the intermonolayer distance, $q\simeq 1/L=0.05$ a.u, electrons feel the interaction with 
other layers. Therefore, the plasmons show a three dimensional character at very small momentum, which is reflected 
in the gap at $|{\bf{q}}|\rightarrow 0$ that can be seen in Fig. \ref{fig:eq_dispersion}. 

\begin{figure}[h!]
\includegraphics[width=\linewidth,clip=true]{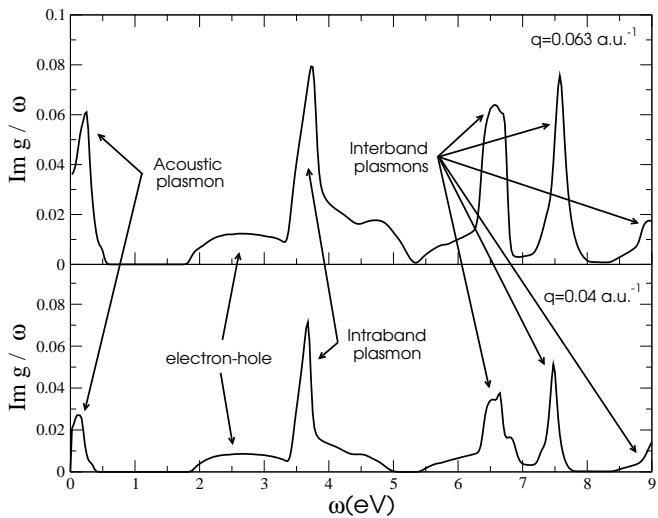}
\caption{Loss function Im $g({\bf{q}};\omega)$ of the hexagonal lithium ML with $r_{s}=2.4$ for two different values of the wavefunction in the $\Gamma K$ direction. Besides the common intraband plasmon, close to the same energy as at equilibrium, and interband one, located at a higher energy, one can also notice a remarkable peak at small $\omega$, the characteristic feature of an acoustic plasmon.}
\label{fig:rs2.4_plasmon}
\end{figure}

\begin{figure}[h!]
\vspace*{0.5cm}
\hspace*{0.8cm}\includegraphics[scale=0.5]{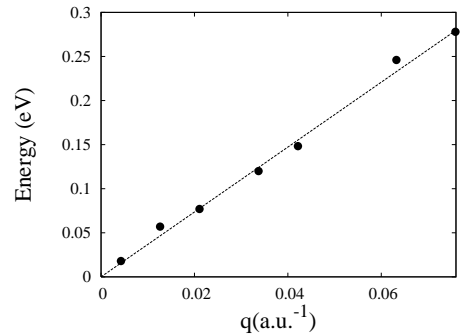}
\caption{Dispersion curve of the \textit{ab initio} acoustic plasmon induced by the applied pressure in a hexagonal lithium  ML with $r_{s}=2.4$. A clear lineal dependence with $|\bf{q}|$ can be observed. Dots are calculations whereas dashed line is a fit. 
The slope of the fit is close to the velocity of the slow electrons. }
\label{fig:rs2.4_dispersion}
\end{figure}

Loss function of the ML with 
$r_{s}=2.4$ at selected momenta in the $\Gamma K$ direction is presented in 
Fig. \ref{fig:rs2.4_plasmon}.
The $s$-bandwidth at this density is very similar to that at equilibrium. As a consequence, the collective excitations
 inside the $s$-band require a very similar energy at both densities, equilibrium and $r_{s}=2.4$. Thus, 
the intraband 
plasmon peak is situated at $\sim$3.5 eV, close to the location at equilibrium. However, the energy gap between the first 
two bands, $s$ and $p_{z}$, increases under pressure, therefore requiring more energy the collective motions of electrons 
between these two bands to be excitated, and shifting the corresponding interband plasmon peak to around $\sim$6.5 eV. 
As well as the two plasmon peaks appearing also in absence of applied pressure, here 
one can also observe peaks associated with two other interband collective excitations with energies $\sim$7.7 eV and 
$\sim$9.5 eV, due to the presence of two partly occupied energy bands. In addition, a remarkable peak is induced by 
pressure at energies close to zero. Besides the common intraband and interband collective excitations, the presence of 
two type of carriers at the Fermi level with different velocities allows the existence of another kind of collective 
modes, known as acoustic plasmons because of their linear dispersion in $|\bf{q}|$ \cite{slava_acoustic}, as it can be checked in Fig. 
\ref{fig:rs2.4_dispersion}. The same acoustic plasmon was also predicted for a berilium ML \cite{aitor_plasmon}. Contrary to the case of two dimensional-like plasmons, which are well established experimentally \cite{nagao1,nagao2}, the acoustic 
counterparts have not yet been unambiguously identified. It is known that, in this case, the light electrons screen the 
Coulomb repulsion between the electrons with a lower velocity, so that an additional intraband acoustic collective 
excitation is produced.

\end{document}